Phase transition, superstructure and physical properties of $K_2Fe_4Se_5$

Y. J. Song[1], Z. Wang[1], Z. W. Wang[1], H. L. Shi[1], Z. Chen[1], H. F. Tian[1], G. F. Chen[2], H. X. Yang[1] and J. Q. Li[1]*

[1] Beijing National Laboratory for Condensed Matter Physics, Institute of Physics, Chinese Academy of Sciences, Beijing 100190, P. R. China

[2] Department of Physics, Renmin University of China, Beijing 100872, P. R. China

The structural features and physical properties of the antiferromagnetic $K_{0.8}Fe_{1.6}Se_2$ (so called $K_2Fe_4Se_5$ phase) have been studied in the temperature range from 300K up to 600K. Resistivity measurements on both single crystal and polycrystalline samples reveal a semiconducting behavior. Structural investigations of $K_2Fe_4Se_5$ by means of transmission electron microscopy and powder x-ray diffraction demonstrate the presence of a well-defined superstructure within the *a-b* plane originating from a Fe-vacancy order along the [130] direction. Moreover, in-situ heating structural analysis shows that $K_{0.8}Fe_{1.6}Se_2$ undergoes a transition of the Fe-vacancy order to disorder at about 600K. The phase separation and the Fe-vacancy ordering in the superconducting materials of $K_xFe_{2-y}Se_2$ (0.2 ⩽ y ⩽ 0.3) has been briefly discussed.

PACS number: 74.62. Bf, 74.60.-w, 74.62.-c, 61.14.Lj, 61.16.Bg

The newly discovered superconductivity in patassium intercalated FeSe [1] and other iron chalcogenide systems [2,3] has triggered a great interest for the investigation of this kind of layered materials and superconducting mechanism [4,5]. Recent studies on these materials reveal some novel phenomena: large moments in antiferromagnetic (AFM) arrangement [6] and resistance hump which is sensitive to the Fe-content [7] and pressure as well [8]. Measurements of neutron diffraction and NMR [9] have focused on the co-existence of magnetic order with superconductivity. Optical investigation of $K_xFe_{2-y}Se_2$ superconducting samples demonstrated the existence of a small energy gap in the superconducting state [10]. It is also noted that these compounds have a rich variety of structural features [11] and magnetic properties [12]. $K_{0.8}Fe_{1.6}Se_2$ (i.e. $K_2Fe_4Se_5$ phase) contains the Fe-vacancy order in association with an AFM structure with $T_N$ ~559K [6,13], accompany with the AFM transition, structure transition and resistivity anomaly was observed [14]. This semiconducting phase has an energy gap of 594 meV as calculated by first-principles calculations [15]. In order to directly reveal the structural changes associated with the phase transition, we carried out a series of *in-situ* transmission electron microscopy (TEM) observations and x-ray diffraction on $K_{0.8}Fe_{1.6}Se_2$ samples from room temperature up to 600 K. We also characterize the superstructure in correlation with the Fe-vacancy order by using powder x-ray diffraction.

Both single crystal and polycrystalline samples of $K_{0.8}Fe_{1.6}Se_2$ were used in our study. The polycrystalline samples were prepared in a two-step solid reaction method: Fe，Se powders and K pieces were firstly placed in a small alumina crucible, and then sealed in a silica tube, each tube filled with 1/3 argon gas. The tube was then preheated at 200 °C for 24 h and then hold at 500℃ for 24 h. The obtained products were then reground, pelleted, and heated at 750℃ for 48 h. The single crystals used in present study were synthesized by Birdgeman method as reported in ref.[7]. Powder x-ray diffraction is performed on a Bruker AXS D8 Advanced diffractometer equipped with TTX 450 holder working in the temperature rang of 80 K ~ 700 K. The temperature dependences of resistivity were measured by a standard four-probe method. Specimens for TEM observations were prepared by peeling off a very thin sheet of a thickness around several tens microns from the single crystal and milling by Ar ion under low temperature. Microstructure and phase transition investigations were performed on a FEI Tecnai-F20 TEM equipped with a double-tilt heating holder.

As an AFM semiconductor, the $K_{0.8}Fe_{1.6}Se_2$ samples often have a large resistivity at low temperatures and the temperature dependence of resistivity almost exhibits a thermally activated behavior. In Fig. 1(a), we show the temperature dependence of resistivity of a polycrystalline $K_2Fe_4Se_5$ measured at temperature from 290 K down to 40 K, and the inset displays the logarithmic form of $\rho(T)$ curve. It is recognizable that the resistivity increase gradually with the decrease of temperature roughly following with the known thermally activated nature for semiconductors: $\rho = \rho_0 exp(E_a/k_B T)$, where $E_a$ is the activation energy. Fig. 1(b) shows the resistivity curve of a $K_4Fe_4Se_5$ single crystal with current going along the *c*-axis direction and within the *a-b* plane, respectively. The inset of the Fig. 1(b) shows that the logarithmic form of $\rho(T)$ curves is fundamentally in agreement with what observed in the polycrystalline sample. In addition, as for the single crystal sample, the in-plane resistivity is a somewhat larger than that along the *c*-axis direction. This fact suggests that the presence of anisotropic transport properties in present system, as similarly discussed in previous literature [16].

In order to understand the fundamental properties of the crystal structure, in particular, the Fe-vacancy ordering in the $K_2Fe_4Se_5$, we have performed a series of structural investigations by means of the x-ray diffraction and TEM observations. The experimental results demonstrate that the polycrystalline samples and the single crystals have the same average structure, and all main reflection peaks can be indexed with lattice parameters *a*=*b*=8.701 Å and *c*=14.036 Å with a space group of *I 4/m* (No. 87). Moreover, the superstructure reflections can be clearly observed in the powder x-ray diffraction pattern taken from the polycrystalline samples. Figure 2 shows an x-ray diffraction pattern obtained from a well-characterized sample at room temperature, the superstructure reflection peaks corresponding with the Fe-vacancy order are indicated by asterisks, these superstructure features is in good agreement with the calculated data for a Fe-vacancy order going along the [130] direction with a period of $L=5d_{130}$. Analysis of diffraction data and structural refinements were performed by using the commercial software MDI Jade 6.5, and the crystallographic data are summarized in Table 1. Compared with the structural data of $KFe_2Se_2$ as reported by J. G. Guo et al. [1], the four interlayer Fe-Fe bonds length are not equal. This fact suggests that visible local structural distortions appear in the $K_2Fe_4Se_5$ crystal resulting from the Fe deficiency. This structural data are fundamentally in agreement with the results obtained by the neutron diffraction which also suggests a quasi-two-dimensional antiferromagnetic structure in association with Fe-vacancy ordering in $K_2Fe_4Se_5$.

TEM observations on the microstructure feature in the K-Fe-Se superconducting system reveal a rich variety of structural phenomena, such as the Fe-vacancy ordering and phase separation. On the other hand, our TEM investigations on the $K_2Fe_4Se_5$ sample in general show a well-defined superstructure. Figure 3(a) shows the high-resolution TEM image taken from a $K_2Fe_4Se_5$ single-crystalline sample, indicating a tetragonal basic structure of the *a-b* plan. This image taken along the [001] zone-axis direction was obtained from a relative thin region of a crystal. The Fe-vacancies positions are clearly recognizable as bright dots and their arrangement and the distance between neighbor bright dots are conformity with $K_2Fe_4Se_5$ structure as demonstrated in our previous publications [11]. The inset of Fig. 3(a) displays a selected-area electron diffraction pattern along the [001]-zone axis direction showing superstructure feature in the *a-b* plane of $K_2Fe_4Se_5$ phase in consistence with our x-ray diffraction data as discussed in above context. Fig. 3(c) shows a high-resolution TEM image taken along the [1-10] zone-axis for the supercell, illuminating the layered structural feature along the *c*-axis direction by the Fe-vacancy ordering along the axis direction. The inset shows the electron diffraction pattern with sharp superstructure spots. This image in contrast with what we observed in the superconducting samples does not show visible phase separation and structural inhomogeneity which were commonly seen in the superconducting materials as discussed in our previous publications [11].

Very recently, experimental investigations on the $K_xFe_{2-y}Se_2$ superconducting materials suggest that this system has a high antiferomagnetic transitions just above 550K, and resistivity anomalies related to Fe-vacancy ordering also appears at almost the same temperature. Fig. 4(a) (upper) shows schematically the evident changes of resistivity and magnetic susceptibility for a typical $K_xFe_{2-y}Se_2$ sample. In order to observe the microstructure alternations in the high temperature range, we have made a careful examination by means of *in-situ* heating TEM observation on a well-characterized $K_2Fe_4Se_5$ sample in the high temperature range of 300 K to 600 K. Fig. 4(b) shows a series of electron diffraction patterns taken along [001] zone axis in the same region on a $K_{0.8}Fe_{1.6}Se_2$ crystal, illustrating the visible changes of the superstructure with the increase of temperature. A diffraction pattern taken at around 300K (upper in Fig. 4(b)) displays a typical pattern of the *a\*-b\** reciprocal plane for the $K_2Fe_4Se_5$ superstructure. Careful examinations of the diffraction patterns at high temperatures reveal clear changes on both intensity and sharpness of the superstructure reflection spots. The intensity change for the superstructure reflection is illustrated in the low-panel of Fig. 4a, illustrating a clearly correlation between physical properties and superstructure. This high-temperature structural transition can be fundamental characterized as the melting of the Fe-vacancy order, it undergoes a transformation, via an intermediate phase with short rang order (500-550K), towards to a Fe-vacancy disorder phase (above 570K). The middle panel of Fig. 4(b) shows a diffraction pattern at 500K in which the main spots can be assigned to a normal tetragonal structure with space group *I 4/mmm* (unit cell parameters $a=b=3.913$ Å, $c=14.10$ Å) and the diffuse reflections indicate the existence of short rang order which can be attributed to the Fe-vacancy ordered state with relatively short coherent length similar with what happened in $Sr_2Nd_{1-x}Ca_xCu_2O_{5+y}F_{1+\delta}$ [17]. As the temperature increases above 550K, the Fe-vacancy order fades progressively away and disappear completely at about 600K (bottom in Fig. 4(b)). The high-temperature crystal structure with out Fe-vacancy order can be well described as the known 122 structure with a tetragonal symmetry of *I 4/mmm*. These results, in combination with previous studies of the neutron diffraction and transport properties, unambiguously suggest that the Fe-vacancy order in this layered system play a critical role for understanding not only the superstructure but also the antiferromagnetic behavior.

In Fig. 4 (c) we show a series of data taken from a powder x-ray diffraction in the temperature rang of 300K up to 600K for a $K_{0.8}Fe_{1.6}Se_2$ sample, the most striking feature revealed in our x-ray diffraction is the disappearance of the superstructure peaks marked by asterisks at about 600K in agreement with our TEM observations. Moreover, $K_2Fe_4Se_5$ phase is found to be unstable at 600K in the Ar environment, the two small peaks marked by arrows belong to the known $FeSe_{0.88}$ phase [18] with $a = 3.784$ Å, $c = 5.532$ Å, and space group *P 4/nmm*.

Actually, the Fe ions in the anti-ferromagnetic $K_2Fe_4Se_5$ has a $Fe^{2+}$ valence state, recent theoretical and experimental investigations on the $K_xFe_{2-y}Se_2$ superconducting system suggest that charge carrier doping can be successfully produced by varying the Fe-concentration in $K_xFe_{2-y}Se_2$. Moreover, the Fe-vacancy order and related antiferromagnetic structure play a critical role for alternation of the superconductivity in present system. Experimental study on a series of samples with chemical composition deviating from $K_2Fe_4Se_5$ by varying either K- or Fe-concentration could result in the appearance of superconductivity in this system. For instance, clear superconductivity with critical transitions between 27K and 33K often appears in $K_{0.8}Fe_{2-y}Se_2$ samples with $0.2 \leqslant y \leqslant 0.3$ and disappears for high Fe-concentration [7]. In addition to the influence on physical properties, the variation of Fe-concentration could also yield clear changes in microstructure such as phase separation and local structural distortions [19]. Actually, in order to understand the correlation between Fe-vacancy order and superconductivity, we have performed a series of TEM investigations on the $K_xFe_{2-y}Se_2$ samples with different chemical composition. In contrast with the well-defined ordered state within the *a-c* superstructure plane as shown in Fig. 3(c), complex phase separation and structural inhomogeneity appears commonly in the superconducting $K_xFe_{2-y}Se_2$ ($0.2 \leqslant y \leqslant 0.3$). In Fig. 5, we show a TEM image taken on a single crystal with a sharper superconducting transition, in which the Fe-vacancy order is invisible in a large fraction of the crystal. Further study on this kind of domain structure shows clear change with the temperature rise as similar discussed in above text for the in-situ heating TEM observations.

In summery, the $K_xFe_{2-y}Se_2$ compounds show a rich variety of structural phenomena in correlation with the Fe-vacancy ordering and the inhomogeneous structure. The important phase ( $K_2Fe_4Se_5$ ) in $K_xFe_{2-y}Se_2$ system was well characterized by powder x-ray diffraction and TEM observations. Our experimental results demonstrate that the $K_2Fe_4Se_5$ phase undergo a phase transition from the well-defined superstructure, via an intermediate phase with short rang order state (500-550K), towards to a Fe-vacancy disorder phase (above 570K). Chemical composition deviating from $K_2Fe_4Se_5$ yields phase separation and structural inhomogeneity in the superconducting materials.

**Acknowledgments**

This work is supported by the National Science Foundation of China, the Knowledge Innovation Project of the Chinese Academy of Sciences, and the 973 projects of the Ministry of Science and Technology of China.

References.

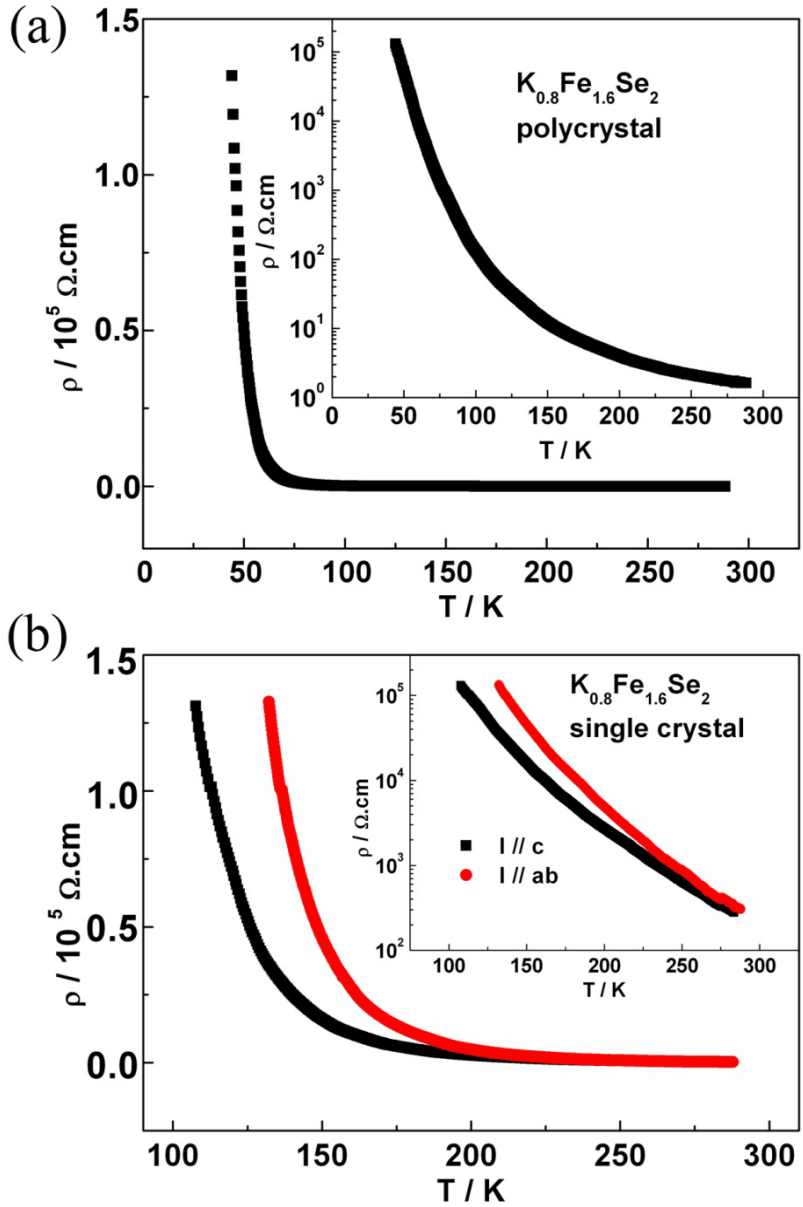

Fig. 1 Temperature dependence of electric resistivity of $K_{0.8}Fe_{1.6}Se_2$: (a) Resistivity curve for a polycrystalline sample and the logarithmic form is shown in the inset. (b) Temperature dependence of resistivity for a $K_{0.8}Fe_{1.6}Se_2$ single crystal with current going along the *c*-axis direction( black ) and within the *a-b* plane( red ), respectively.

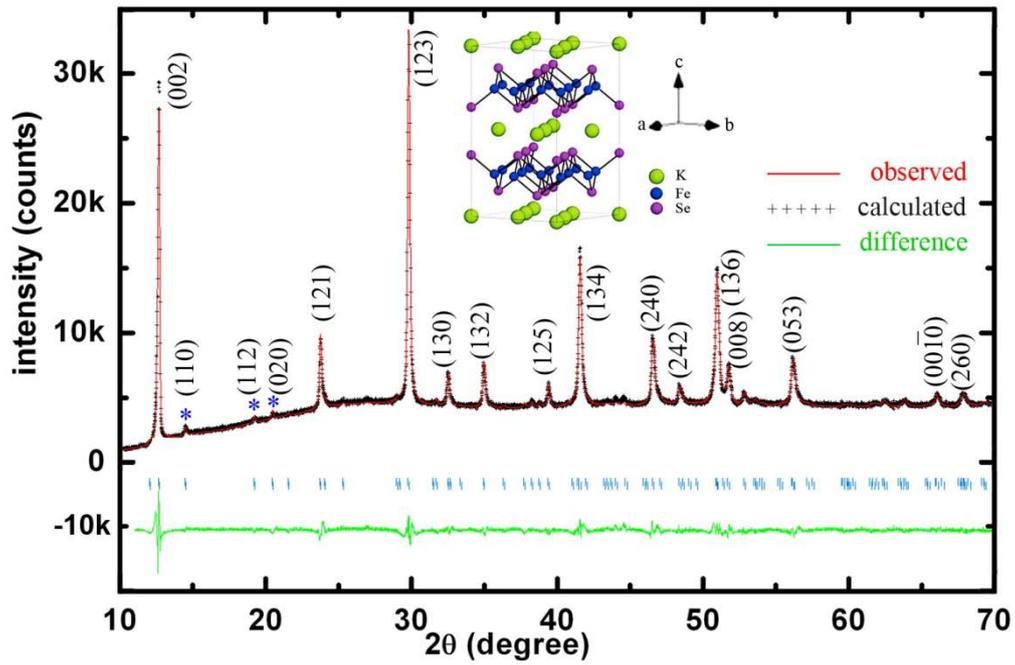

Figure 2 Powder x-ray diffraction pattern taken from a $K_{0.8}Fe_{1.6}Se_2$ polycrystalline sample : the red line is experimental diffraction pattern observed at room temparature and the plus sign is the calculated pattern using the structural data in Table 1. The inset shows a schematic structural model of $K_2Fe_4Se_5$ .

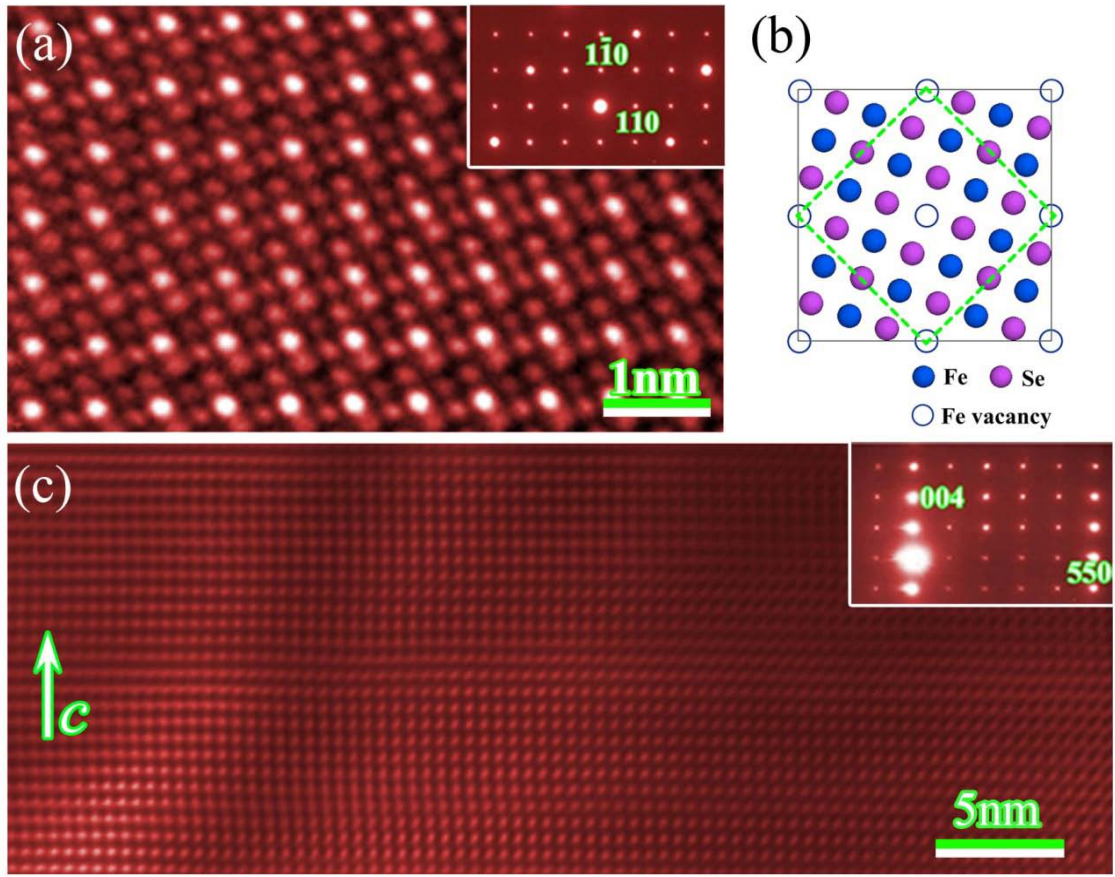

Figure 3 High resolution TEM images taken along (a) the [001] and (c) the [1-10] zone-axis directions reveal the atomic structure features in $K_{0.8}Fe_{1.6}Se_2$ crystals. Insets show the corresponding electron diffraction patterns, (b) Top view of the Fe-Se layer and the dash line marks the superstructure unit cell.

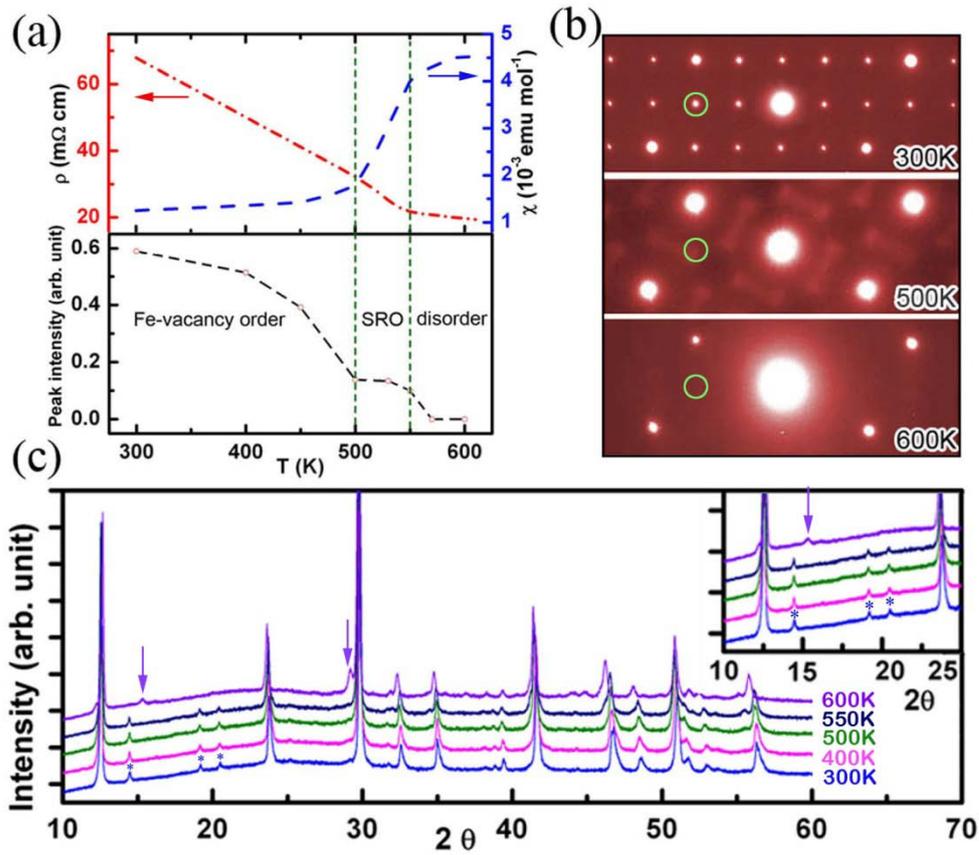

Figure 4 (a) Changes of resistivity and magnetic susceptibility of $K_{0.8}Fe_{1.6}Se_2$ showing the anomalies above 550K. Lower panel shows the alternation of superstructure as observed by TEM. (b) Electron diffraction patterns at different temperature, (c) the XRD patterns of $K_{0.8}Fe_{1.6}Se_2$ at different temperature, the two arrows marks the (001) and (011) peaks of $FeSe_{1-x}$ at around 600K, respectively.

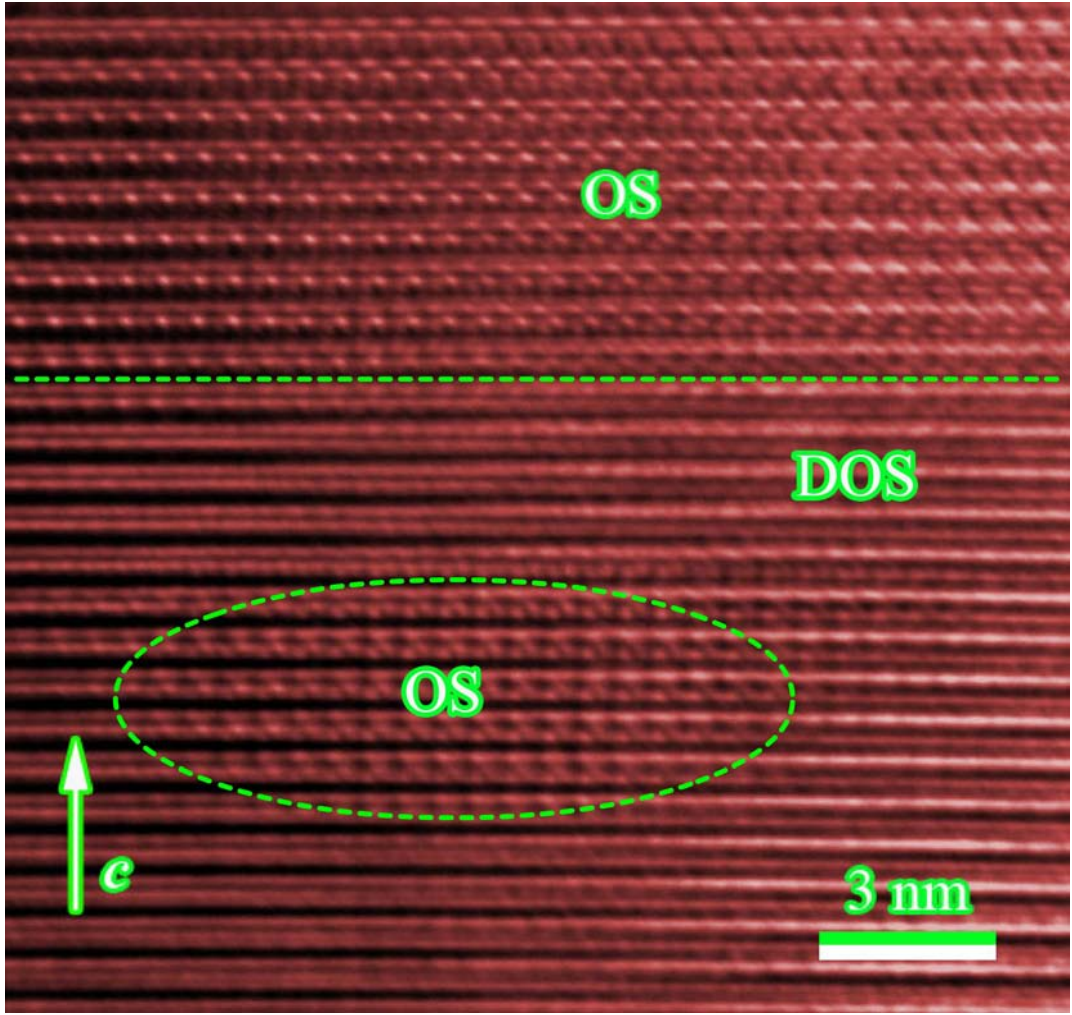

Fig. 5 A high-resolution TEM image taken along [130] zone-axis direction on a single crystal with a sharper superconducting transition, in which both ordered state (OS) and disordered state (DOS) are observed.

Table 1 The structural data for $K_2Fe_4Se_5$ as obtained from the x-ray diffraction.

| Formula | $K_2$ $Fe_4$ $Se_5$ | |
|---|---|---|
| Temperature(K) | 297 | |
| space group | I 4/m (87) | |
| a (Å) | 8.7235（5） | |
| c (Å) | 14.1285（0） | |
| V (Å3) | 1075.183（1） | |
| z | 4 | |
| Rp | 2.06% | |
| Rwp | 2.98% | |
| Rexp | 1.39% | |
| | | |
| Atomic parameters | | occ. |
| K1 | 2a (0, 0, 0) | 1 |
| K2 | 8h (0.36915, 0.15771, 0) | 0.8422 |
| Fe1 | 4d (0, 1/2, 1/4) | 0.3852 |
| Fe2 | 16i(0.1932, 0.1003, 0.2488) | 0.9093 |
| Se1 | 4e (1/2, 1/2, 0.1395) | 1 |
| Se2 | 16i (0.1219, 0.2998, 0.1462) | 1 |